# Nuclear quadrupole interactions of 111In/Cd solute atoms in a series of rare-earth palladium alloys

Qiaoming Wang, Gary S. Collins

*Dept of Physics and Astronomy, Washington State University, Pullman, WA 99164, USA*

+1-509-335-1354

+1-509-335-7816

collins@wsu.edu

http://defects.physics.wsu.edu/

Nuclear quadrupole interactions were measured at $^{111}$In/Cd probe atoms in rare-earth palladium phases $RPd_3$ having the $L1_2$ structure using the technique of perturbed angular correlation of gamma rays (PAC). Measurements were made for pairs of samples having compositions of the Pd-poorer and Pd-rich boundaries of the $RPd_3$ phase fields, typically 75 and 78 at.% Pd. Two signals were detected in most phases, corresponding to probe atoms on cubic R- and non-cubic Pd-sublattices. Site preferences of parent In-probe atoms were characterized by site-fractions of probes on the R- and Pd-sublattices. For all Pd-rich samples, probes exclusively occupied the R-sublattice, consistent with a heuristic rule that solute atoms tend to occupy the sublattice of an element in which there is a deficiency. For Pd-poorer alloys with R= Tb, Er, Yb, Lu, Y and Sc, probes exclusively occupied the Pd-sublattice. For Pd-poorer alloys with R= Ce, Pr, Nd, Sm and Eu, probes occupied both sublattices, with site fractions varying as a function of temperature. In contrast, probes only occupied the R-sublattice in Pd-poorer $LaPd_3$. Jump frequencies on the Pd-sublattice of daughter Cd-probes were determined from nuclear relaxation caused by fluctuating electric field gradients. Activation enthalpies for diffusional jumps were determined from temperature dependences and found to increase linearly as the lattice parameter decreases along the series Pr, Nd, Eu and Sm. Jump frequencies are believed to have been even higher in $CePd_3$ than in $PrPd_3$, but were too low to be detectable in Tb, Er, Yb, Lu, Y and Sc palladides. A correlation between site preferences and jump frequencies is noted and discussed. This paper provides a complete account of measurements of jump frequencies of Cd-probe atoms and comparisons with similar measurements made on other series of $L1_2$ phases.

PAC: perturbed angular correlation of gamma rays; EFG: electric field gradient

## Introduction

Perturbed angular correlation of gamma rays (PAC) can be used to determine frequencies of diffusional jumps in solids through measurements of nuclear quadrupole relaxation. Such relaxation is caused by a change in orientation or magnitude of the electric field gradient (EFG) in each jump [1], as is the case for many diffusion sublattices. Extensive measurements have been made for $^{111}$Cd PAC probes jumping on the A-sublattice in series of rare-earth intermetallic compounds $A_3B$ having the $L1_2$ ($Cu_3Au$) structure, including indides [1, 2, 3, 4], stannides [3, 5], and gallides [6].

The present work extends those jump-frequency measurements to the rare-earth palladides $RPd_3$, which form in $L1_2$ structure with all lanthanide elements, Y and Sc [7]. The phases have narrow field widths [8], typically extending from 75 to 78 at.% Pd. Jump-frequency activation enthalpies were determined from temperature dependences for Pr, Nd, Sm and Nd phases and found to increase linearly with decreasing lattice parameter, starting from a low value for $PrPd_3$. In addition to jump frequencies, fitted site fractions of probe atoms on R and Pd sites were used to determine site preferences of the $^{111}$In parent probe atoms. Signals for probes on the two sites are readily resolved because the R site is cubic and the Pd site is not, leading respectively to zero and nonzero values of the electric field gradient tensors at the probe sites. This paper gives a complete account of the jump frequency measurements and a summary of site-preference results that will be reported more fully elsewhere [9].

## Experimental methods

Rare-earth palladium alloys were made by arc-melting high-purity metals together with $^{111}$In activity under argon. Samples included $RPd_3$ phases with rare-earth elements R= La, Ce, Pr, Nd, Sm, Eu, Tb, Er, Yb, Lu, Y and Sc. Binary alloy phase diagrams show that the $L1_2$ phase fields typically extend from 75 to 78 at.% Pd [8]. Previous work on other intermetallic compounds in this laboratory has shown that both jump frequencies and site preferences are highly sensitive to precise alloy compositions, even for “line compounds” that have unmeasured, small field widths [10]. Accordingly, to maintain a constant composition in the face of possible changes in composition due to evaporation of one or other alloy constituent element during measurement at high temperature, samples were made in pairs having average compositions of approximately 73 and 80 at.% Pd, leading to two-phase mixtures having a large volume fraction of the desired $L1_2$ phase, a small volume fraction of an adjacent phase that could be accounted for in the fitting, as necessary. This methodology leads to well defined boundary compositions for the samples that are labeled below as “Pd-richer” or “Pd-poorer”. This terminology is preferred, for example, over “Pd-poor” because a Pd-poorer sample might in fact be Pd-rich with respect to the stoichiometric composition, in light of uncertainties in boundary compositions taken from binary phase diagrams.

PAC measurements were made using four-counter spectrometers of standard design (see refs. [11, 12] for further details). Spectra were fitted with a superposition of two axially-symmetric perturbation functions for quadrupole interaction, $G_2(t)$, one having zero quadrupole interaction frequency and the other having non-zero frequency, as expected for probe atoms in the $RPd_3$ phase on cubic R-sites and non-cubic Pd-sites. The general form of the perturbation function used for each site in the fitting was

$$G_2(t) \cong \begin{pmatrix} \frac{1}{5} + \frac{13}{35}\cos\omega_1 t \exp(-\frac{(\sigma t)^2}{2}) + \frac{10}{35}\cos 2\omega_1 t \exp(-\frac{(2\sigma t)^2}{2}) + \\ + \frac{5}{35}\cos 3\omega_1 t \exp(-\frac{(3\sigma t)^2}{2}) \end{pmatrix} \exp(-\lambda t), \quad (1)$$

in which $\omega_1 = \frac{3\pi}{10}\frac{eQV_{zz}}{h}$ is the fundamental observed frequency, defined in terms of the nuclear quadrupole moment $Q$ and principal component of the EFG tensor, $V_{zz}$, $\sigma$ represents inhomogeneous frequency broadening caused by weak, static EFG’s produced by point defects, and $\lambda$ is a dynamical relaxation frequency. Eq. 1 is the product of a static perturbation function and a dynamical relaxation factor $\exp(-\lambda t)$ and applies in the slow-fluctuation dynamical regime, when the jump-frequency $w < \omega_1$. Eq. 1 is a general form frequently used for fitting nuclear relaxation. For diffusion of probes on the Pd-sublattice of the $L1_2$ structure, it has been shown that the relaxation frequency $\lambda$ is equal in very good approximation to $w$, the mean jump frequency (that is, inverse of the mean residence time) [1]. The total perturbation function is the sum of two functions of the form of Eq. 1, each multiplied by a signal amplitude equal to the site-fraction for that site. Site fractions for probes on the R- and Pd-sublattices were fitted and used to evaluate site preferences that will be presented and analyzed in detail elsewhere [9]. Magnitudes and temperature dependences of the quadrupole interaction frequencies will also be presented and discussed elsewhere [9].

## Results

We consider in turn results for palladium alloys with different groups of rare-earth elements.

### Heavy lanthanides Tb, Er, Yb, Lu

Fig. 1 shows representative PAC spectra for $LuPd_3$ for Pd-richer (A) and Pd-poorer (B) phase boundary compositions on the left and right, respectively. Spectra measured for the Pd-poorer composition (B) exhibit a high-frequency, ~100 Mrad/s, axially symmetric quadrupole perturbation attributed to probes on Pd-sites. For the Pd-richer sample (A), spectra show a low-frequency interaction that is attributed to probe atoms on R-sites. Since the boundary composition is ~78 at.% Pd, it is not a surprise that there is inhomogeneous broadening for the Pd-richer sample because compositional point defects must be present in order to accommodate the deviation from stoichiometry. The most likely such defects are Pd antisite atoms, $Pd_R$, For a composition of 78 at.% Pd, the mole fraction of antisite atoms on the R-sublattice would be 12%. The inhomogeneous broadening is very small for such a large mole fraction of point defects. This is because $Pd_R$ defects will be in the second or further neighbor shells of probe atoms on the R-sublattice, $In_R$, producing only weak EFGs. The difference in site preference of indium solute atoms at the two boundary compositions shown in Fig. 1 is consistent with the heuristic rule that solute atoms tend to occupy the sublattice of a host element in which there is a deficiency [13]. This rule follows from the general expectation that the energy of an ordered alloy is smaller when the overall number of point defects is reduced. Putting the solute atom on the sublattice of an element for which there is an excess would increase the defect count by one. Spectra for the Pd-poorer alloys exhibited only a small or zero site fraction for probes on R-sites and there was no change in the site fractions with increasing temperature. For Pd-poorer alloys (B), it can be seen that there is a small amount of inhomogeneous broadening in the spectrum at room temperature (RT), where there can be negligible diffusional broadening due to atomic motion. However, no increase in damping is observed at higher temperature that could be attributed to diffusional motion. It is concluded that the jump frequency of indium solutes in $LuPd_3$ is less than ~1 MHz at the highest temperature measured, 1198 K. Similar results were observed for all Pd-poorer samples formed with the heavy lanthanide elements Tb, Er, Yb and Lu, and also with Y and Sc. In summary, jump frequencies were too small to be detected for this group of rare earths. Site-preferences were very strong for either R- or Pd-sites, depending on the boundary composition, and were independent of temperature. While jump frequencies must be greater than ~1 MHz in order to be detected in PAC spectra, site-fractions should equilibrate quickly during the one-day duration of measurement if jump frequencies are of order $10^{-8}$ MHz or greater. Thus, thermodynamic equilibrium between solute site occupations is readily achieved even when there is no detectable evidence for diffusional jumps.

**Intermediate lanthanides Sm, Eu**

Spectra for Pd-richer $SmPd_3$ and $EuPd_3$ exhibited only low frequency signals for probes on R-sites, as in Fig. 1(A). Spectra for Pd-poorer alloys exhibited changes with temperature that are illustrated in Fig. 2 for $EuPd_3$. At low temperature (673K), the spectrum is dominated by a high frequency ~100 Mrad/s signal that is attributed to probe atoms on the Pd-sublattice. With increasing temperature, the site fraction for the low-frequency signal attributed to probes on the R-sublattice increases at the expense of the Pd-sublattice signal. Therefore, one can conclude that Pd-sites are preferred at low temperature and that R-site occupation is thermally activated at higher temperatures. In addition to a change in signal amplitude, the Pd-site signal becomes increasingly damped due to diffusional relaxation at higher temperatures. Each time spectra was fitted using Eq. 1 to obtain a table of jump frequencies $\lambda = w$ as a function of temperature. The jump frequencies were then fitted to a thermally activated expression of the form

$$w = w_0 \exp(-Q / k_B T), \qquad (2)$$

in order to determine the jump-attempt frequency prefactor $w_0$ and jump-frequency activation enthalpy $Q$.

## Lighter lanthanides Pr, Nd

Just as for the intermediate lanthanides, it was found that probe atoms distribute on both R- and Pd- sublattices and that jump frequencies were measurable and increased with temperature. Fig. 3 shows representative spectra for a Pd-richer sample of $PrPd_3$, which, like in Fig. 1(A), demonstrate a strong site preference of indium solutes for the cubic R-sublattice. There is clearly less inhomogeneous broadening in Fig. 3 than in Fig. 1(A), suggesting that the boundary composition in Pd-richer $PrPd_3$ is closer to the stoichiometric 75 at.% composition than in $LuPd_3$, leading to a smaller concentration of constitutional point defects such as $Pd_R$ antisite atoms.

Fig. 4 shows measurements for Pd-poorer $PrPd_3$, with time-domain spectra on the left and corresponding frequency spectra on the right. Similar to Fig. 2, the time-series spectra in Fig. 4 exhibit increased damping with increasing temperature that is well described by the exponential damping factor in Eq. 1. Frequency spectra show equal dynamical broadening of the three frequency components of the static perturbation function. Fig. 5 shows an Arrhenius plot of fitted jump frequencies for $PrPd_3$ spectra, yielding $w_0$= 0.4(1.1) THz and $Q$= 0.77(18) eV for the fit shown by the drawn line. (The two low temperature points in the figure were assumed to have been disturbed by a small amount of static, inhomogeneous broadening and were not included in the fit.)

The most likely diffusion mechanism for probe atoms on the Pd-sublattice in $RPd_3$ phases is the simple Pd-sublattice vacancy diffusion mechanism, in which Pd-vacancies exchange with probe atoms. For this mechanism, no atomic disorder is created in each jump. For vacancy diffusion, the jump-attempt prefactor is generally expected to be of the order of the vibrational frequency of an atom in the solid, or about 1 THz. The fitted value of 0.4 THz is therefore entirely reasonable in magnitude. Similar analysis was carried out for all four of these intermediate phases, with results tabulated and discussed below.

Visual inspection of Fig. 4 shows that the site preference of indium solutes is temperature dependent. The site fraction on R-sites is evident in amplitudes of low-frequency components of the spectra, which are large at lower temperature and decrease with increasing temperature. The complementary site fraction of the (high-frequency) signal for probe atoms on Pd-sites can be visualized most readily via the height of the central peak at *t*=0, which grows with increasing temperature. Thus, the preferred site of indium solutes in $PrPd_3$ at low temperature is the R-site, and occupation of Pd-sites is thermally activated and increases with increasing temperature, opposite to the behavior shown in Fig. 2. Thus, the ground-state site preference of indium solutes has changed from Pd-sites in Sm and Eu phases to R-sites in Pr and Nd phases.

### Lanthanum Palladide

Unlike for all other rare-earth palladium phases studied, spectra for Pd-poorer $LaPd_3$ exhibited only the low-frequency signal attributed to indium on R-sites (see Fig. 6). No high-frequency signal was detected that could be attributed to solutes at Pd-sites. Thus, the R-site appears to be so highly preferred that no thermal activation of solutes to Pd-sites was observed with increasing temperature (cf. Fig. 6). As an alternative explanation, the boundary composition for Pd-poorer $LaPd_3$ might actually be slightly Pd-rich, but close examination of phase diagrams for the lanthanide palladides [8] did not suggest any systematic trend that would set $LaPd_3$ apart.

Interestingly, spectra in Fig. 6 exhibit a *decrease* in broadening at high temperature. As in other spectra measured for Pd-richer $RPd_3$ phases (cf. Figs. 1 and 3), the R-site signal in Pd-poorer $LaPd_3$ exhibited inhomogeneous broadening near room temperature (RT) that is attributed to weak EFG disturbances caused by point defects due to an off-stoichiometric composition. However, it can be seen in the figure that the observed broadening decreases at higher temperatures, nearly disappearing in the spectrum measured at 873 K. This effect was reversible with temperature. As a possible explanation, the decrease in broadening is attributed to increasingly rapid motion of defects near the probe atoms, leading to motional averaging of the weak EFGs. As discussed below, jump frequencies at a given temperature on the Pd-sublattice increase in sequence in the Pd-poorer palladides formed with Sm, Eu, Nd, Pr and (possibly) Ce, so it should not be surprising that jump frequencies might be even greater in $LaPd_3$. However, the magnitude of the jump frequencies of vacancies near the In-probe atom is difficult to estimate. since indium probes are excluded from the Pd-sublattice due to a strong site preference. Analysis would require a different diffusion model to describe the weak EFGs and jump transitions that is beyond the scope of this paper.

Evidence of another phase was observed in spectra of Pd-richer $LaPd_3$ (not shown). Measurements exhibited a 40% site fraction of a well-defined quadrupole interaction signal with $\omega_1$= 129(2) Mrad/s and EFG asymmetry parameter $\eta$= 0.49(1). Since no intermediate phase was reported to exist between the $LaPd_3$ phase and Pd-metal [8], x-ray analysis was carried out that confirmed presence of large signals from the expected $L1_2$ phase and from $LaPd_5$ ($CaCu_5$ structure, $D2_d$, [14]). There are two inequivalent Pd-sites in the phase, one axially symmetric and the other non-axially symmetric. Since the EFG asymmetry parameter was not zero, as it would be for axial symmetry, indium apparently prefers the latter site strongly.

### Cerium palladide

PAC spectra measured for Pd-poorer $CePd_3$ are shown in Fig. 7. The site fraction for probe atoms on the R-sublattice is much larger at low temperature than observed for $PrPd_3$ (see Fig. 4). The site fraction for probes on the Pd-sublattice increases rapidly with increasing temperature, which is most readily visible in Fig. 7 in the rapidly growing amplitude of the central peak at $t$=0. At the same time, increasing diffusional relaxation leads to damping of the high-frequency signal. However, spectra at 873 and 923K exhibit a central peak in the time domain that is twice broader than expected for a static quadrupole interaction signal with $\omega_1 \approx 100$ Mrad/s and, at the same time, a small amplitude of the ~100 Mrad/s signal identified with the expected $In_{Pd}$ probes persists. This behavior appears to arise from two ensembles of probes on Pd-sites that experience different jump frequencies. While general features of the two spectra at the highest temperatures, 873 and 923K, are consistent with high jump frequencies, the broadened central peak and persistence of a small site fraction for the high-frequency signal cannot be reconciled with the simple model used to interpret observations for Pr, Nd, Sm and $EuPd_3$. Accordingly, estimates are not given for jump frequencies in $CePd_3$ although they appear to be very high.

The temperature dependence of the site-fraction ratio could be fitted well, taking the sum of the amplitudes of the high-frequency signal and central peak to be the site fraction of probes starting on Pd-sites and the sum of low-frequency signals to be the fraction of probes starting on R-sites.

Site preferences for $CePd_3$ appear to be transitional between those for $LaPd_3$ and $PrPd_3$. The R-site is the preferred low-temperature site for indium solutes in $LaPd_3$, $CePd_3$, $PrPd_3$, and also $NdPd_3$. For $SmPd_3$ and $EuPd_3$, on the other hand, the preferred site at low temperature is the Pd-site, with R-site occupation being thermally activated. For the heavier lanthanides, Tb to Lu, the ground-state site in Pd-poorer alloys is the Pd-site, and no thermal activation was observed to the R-site. It is interesting that the site preference switches between $NdPd_3$ and $SmPd_3$. It is not known why relative energies of indium solutes on R-sites and Pd-sites change along the lanthanide series.

### Jump frequencies along the series of lanthanide palladides

Arrhenius plots of jump-frequencies on the Pd-sublattice are collected together in Fig. 8 for Pr-, Nd-, Eu- and Sm-$Pd_3$. Jump frequencies decrease in order from $PrPd_3$ to $SmPd_3$. Results of fits using Eq. 2 are given in Table 1 for the four phases. Activation enthalpies can be seen to increase from 0.77 eV for $PrPd_3$ to 2.2 eV for $SmPd_3$. Fig. 9 shows a plot of the fitted activation enthalpies as a function of lattice parameter. It can be seen that the activation enthalpies vary linearly with lattice parameter. The reason for this empirical dependence on lattice parameter is not known.

In the palladides, one observes diffusion of probes that start on Pd-sites via damping of the high-frequency signal. Since the Pd-site is not the preferred low-temperature site in Ce, Pr and $NdPd_3$ phases, the ensemble of probes that start on Pd-sites at time $t$=0 can be thought of as energetically “unstable” (and most unstable for $CePd_3$ and $LaPd_3$, which have the largest ratios of $f_R / f_{Pd}$ ). That instability appears to promote higher jump frequencies in those three phases. Thus, there is a coupling between site preferences and jump frequencies that helps to explain why jump frequencies in Fig. 8 are greatest for $PrPd_3$. Elsewhere, a quantitative analysis of the temperature dependences of site-fraction ratios will be presented and related more fully to the jump frequency behavior reported here [9].

## Jump frequencies for palladides compared with those for indides and other $L1_2$ series

Arrhenius plots of jump frequencies have been measured in this laboratory for over 40 compositions in 30 phases. To provide an overview for such a large collection of measurements, it is useful to characterize each Arrhenius plot using a single parameter rather than the two parameters $w_0$ and $Q$. This will be taken to be the temperature $T_{10}$ at which the jump frequency is equal to 10 MHz, an approach first applied for the indides [4] and later extended to include stannides, aluminides and gallides [5]. For example, from Fig. 5 it can be seen that the jump frequency of Cd on the Pd-sublattice in $PrPd_3$ is 10 MHz at reciprocal temperature $1/k_BT_{10} = 13.8 eV^{-1}$. Fig. 10 summarizes data for all systems studied, with the $1/k_BT_{10}$ scale shown on the left and an ordinary temperature scale on the right. Jump frequencies are generally greater for phases having greater values of $1/k_BT_{10}$. It can be seen that the indide and palladide series show similar strong increases in jump frequency with increasing lattice parameter among the lightest lanthanide phases. A possible explanation for these trends is that vacancy formation enthalpies may be smaller for the light lanthanide phases. This would lead to greater deviations of the boundary compositions from the stoichiometric composition in the lighter lanthanide phases, including a greater concentration of structural vacancies at the boundary composition that would assist diffusion.

While lattice parameters in Fig. 10 span a large range, from 0.40-0.48 nm, no universal dependence of jump frequencies on lattice parameter is evident. Considering that the upper temperature limit of spectrometers in these experiments was ~1300K, non-observation of nuclear relaxation on the Pd-sublattice in Tb, Er, Yb, Lu, Y and Sc palladides can simply be attributed to values of $1/k_BT_{10}$ that are below about 8 $eV^{-1}$, making relaxation undetectable. This is suggested by the dashed line in Fig. 10 that extends the trend in the series of lanthanide palladides to still smaller lattice parameters.

## Summary

PAC measurements of nuclear quadrupole interactions in rare-earth palladium phases of $L1_2$ structure were used to determine site preferences of parent In-probe atoms and jump frequencies of daughter Cd-probe atoms. Measurements were made for pairs of alloys having opposing boundary compositions of the $L1_2$ phase field. Probes had strong site preferences for rare-earth sites in Pd-richer alloys. In Pd-poorer alloys, probes had strong site preferences for Pd-sites in the heavier lanthanide phases (Tb to Lu, and Y and Sc), intermediate site preferences with temperature dependent site fractions for lighter lanthanide phases (Ce to Eu), and a strong preference for only the La-site in $LaPd_3$. Nuclear relaxation was observed for probes starting on the Pd-sublattices at time $t$=0 in Ce. Pr, Nd, Sm and Eu phases. Jump frequencies were fitted and activation enthalpies were determined for Pr to Eu phases. A correlation was noted between site-preferences and jump frequencies, with jump frequencies increasing as the site-preference of indium shifted to the R-site from the Pd-site. Finally, jump frequency trends for lanthanide palladides were compared with those for other series of $L1_2$ phases.

Supported in part by the National Science Foundation under grant NSF DMR 90-04096. Randal Newhouse is thanked for assistance.

## References

1. Zacate, M.O., Favrot, A., Collins, G.S.: Atom Movement in $In_3La$ studied via Nuclear Quadrupole Relaxation , Physical Review Letters 92, 225901(1-4) (2004); and *Erratum*, Physical Review Letters 93, 49903(1) (2004)

2. Collins, G.S., Favrot, A., Kang, L., Solodovnikov, D., Zacate, M.O.: Diffusion in intermetallic compounds studied using nuclear quadrupole relaxation, Defect and Diffusion Forum 237-240, 195-200 (2005)

3. Collins, G.S., Favrot, A., Kang, L., Nieuwenhuis, E.R., Solodovnikov, D., Wang, J. and Zacate, M.O.: PAC probes as diffusion tracers in solids, Hyperfine Interactions 159, 1-8 (2005)

4. Collins, G.S., Jiang, X., Bevington, J.P., Selim, F., Zacate, M.O.: Change in diffusion mechanism with lattice parameter in the series of lanthanide indides having $L1_2$ structure, Phys. Rev. Lett. 102, 155901 (2009)

5. Harberts, M.L., Norman, B., Newhouse, R., Collins, G.S.: Comparison of jump frequencies of $^{111}$In/Cd tracer atoms in $Sn_3R$ and $In_3R$ phases having the $L1_2$ structure (R= rare earth), Defect and Diffusion Forum, vol. 311, p. 159 (2011)

6. Jiang, X., Zacate ,M.O., Collins, G.S.: Jump frequencies of Cd tracer atoms in $L1_2$ lanthanide gallides, Defect and Diffusion Forum 289-292, 725 (2009).

7. Wang, Qiaoming: Site preferences and jump frequencies of In/Cd in rare-earth palladium phases having the $L1_2$ structure, MS Thesis, Washington State University, May 2012 (unpublished).

8. Massalski, T.B., Okamoto, H., Subramanian, P.R., Kacprzak, L.: Binary alloy phase diagrams, ASM, Materials Park, Ohio, second ed. (1990)

9. Wang, Q., Collins, G.S. (to be published)

10. Collins, G.S.: Nonstoichiometry in line compounds, Journal of Materials Science 42, 1915-1919 (2007)

11. Schatz, G., Weidinger, A.: Nuclear Condensed Matter Physics, John Wiley, New York (1996)

12. Zacate, M.O., Collins, G.S.: Temperature- and composition-driven changes in site occupation of solutes in $Gd_{1+3x}Al_{2-3x}$, Physical Review B69, 174202(1-9) (2004)

13. Zacate, M.O., Collins, G.S.: Composition-driven changes in lattice sites occupied by indium solutes in $Ni_2Al_3$ phases, Physical Review B70, 24202(1-17) (2004)

14. Ning Y.T.et al.: J. of the Less Common Metals 147, 167-173 (1989)

## Figure Captions

Fig 1. PAC spectra of $LuPd_3$ at the opposing boundary compositions A (Pd-richer) and B (Pd-poorer). Double-sided PAC spectra such as these show data at negative times that mirror the data at positive coincidence times. They are intended to make it easier to visualize the nuclear relaxation caused by diffusion.

Fig. 2. PAC spectra for Pd-poorer $EuPd_3$. With increasing temperature, a shift in site preference can be seen from the high-frequency Pd-site to the low-frequency R-sites. Diffusional damping of the high-frequency signal increases with temperature.

Fig 3. PAC spectra of $PrPd_3$ for the Pd-richer boundary composition. The low frequency interaction is attributed to indium probe atoms on the cubic Pr-sublattice.

Fig 4. PAC spectra for $PrPd_3$ at the Pd-poorer boundary composition. Time domain spectra are on the left and fourier transforms on the right. Diffusional relaxation is observable as damping in the time-domain or frequency broadening in the frequency spectra.

Fig 5. Arrhenius plot of jump frequencies of indium probes in a sample of Pd-poorer $PrPd_3$. The two lowest temperature points were excluded from the fit shown by the drawn line.

Fig 6. PAC spectra for Pd-poorer $LaPd_3$, exhibiting only the signal for indium solutes on R-sites.

Fig 7. PAC spectra for Pd-poorer $CePd_3$ measured at the indicated temperatures. With increasing temperature, there is a rapid shift in site preference from the Ce-site (low frequency) to the Pd-site (high frequency) and, at the same time, an increase in diffusional nuclear relaxation of the high-frequency signal. This combination of changes leads to complex spectra..

Fig 8. Arrhenius plots of jump frequencies of Cd probe atoms on the Pd-sublattices in the indicated $L1_2$ phases.

Fig 9. Activation enthalpies for diffusional jumps of Cd probe atoms on the Pd-sublattices of four $RPd_3$ phases plotted versus lattice parameter.

Fig.10. Inverse temperatures at which jump frequencies of Cd probe atoms equal 10 MHz in $L1_2$ phases of rare-earth elements with In, Sn, Ga, Al and Pd. “A” and “B” designate R-poorer and R-richer boundary compositions.

Table 1. Activation enthalpies and prefactors for diffusional jumps of Cd solute atoms on the Pd-sublattice. Also shown are lattice parameters of the phases.

| Phases | $Q$ (eV) | $w_0$ (THz) | Lattice param. (nm) |
|---|---|---|---|
| $CePd_3$ | - | - | 0.4160 |
| $PrPd_3$ | 0.77(18) | 0.4(11) | 0.4135 |
| $NdPd_3$ | 1.10(28) | 6(21) | 0.4120 |
| $EuPd_3$ | 1.14(11) | 3(3) | 0.4101 |
| $SmPd_3$ | 2.19(13) | $3(4)\times10^3$ | 0.4068 |

Fig 1

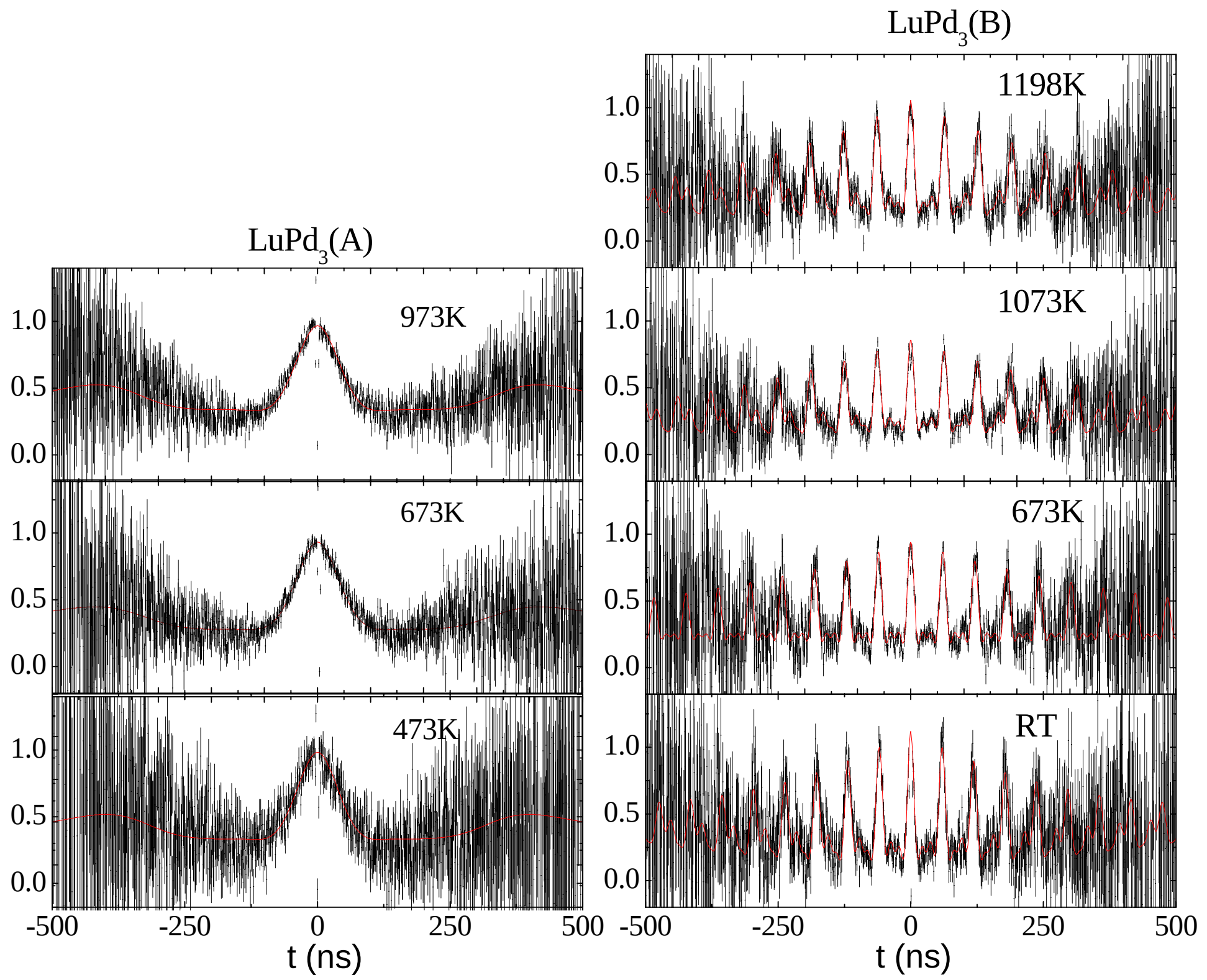

Fig 1. PAC spectra of $LuPd_3$ at the opposing boundary compositions A (Pd-richer) and B (Pd-poorer). Double-sided PAC spectra such as these show data at negative times that mirror the data at positive coincidence times. They are intended to make it easier to visualize the nuclear relaxation caused by diffusion.

Fig 2

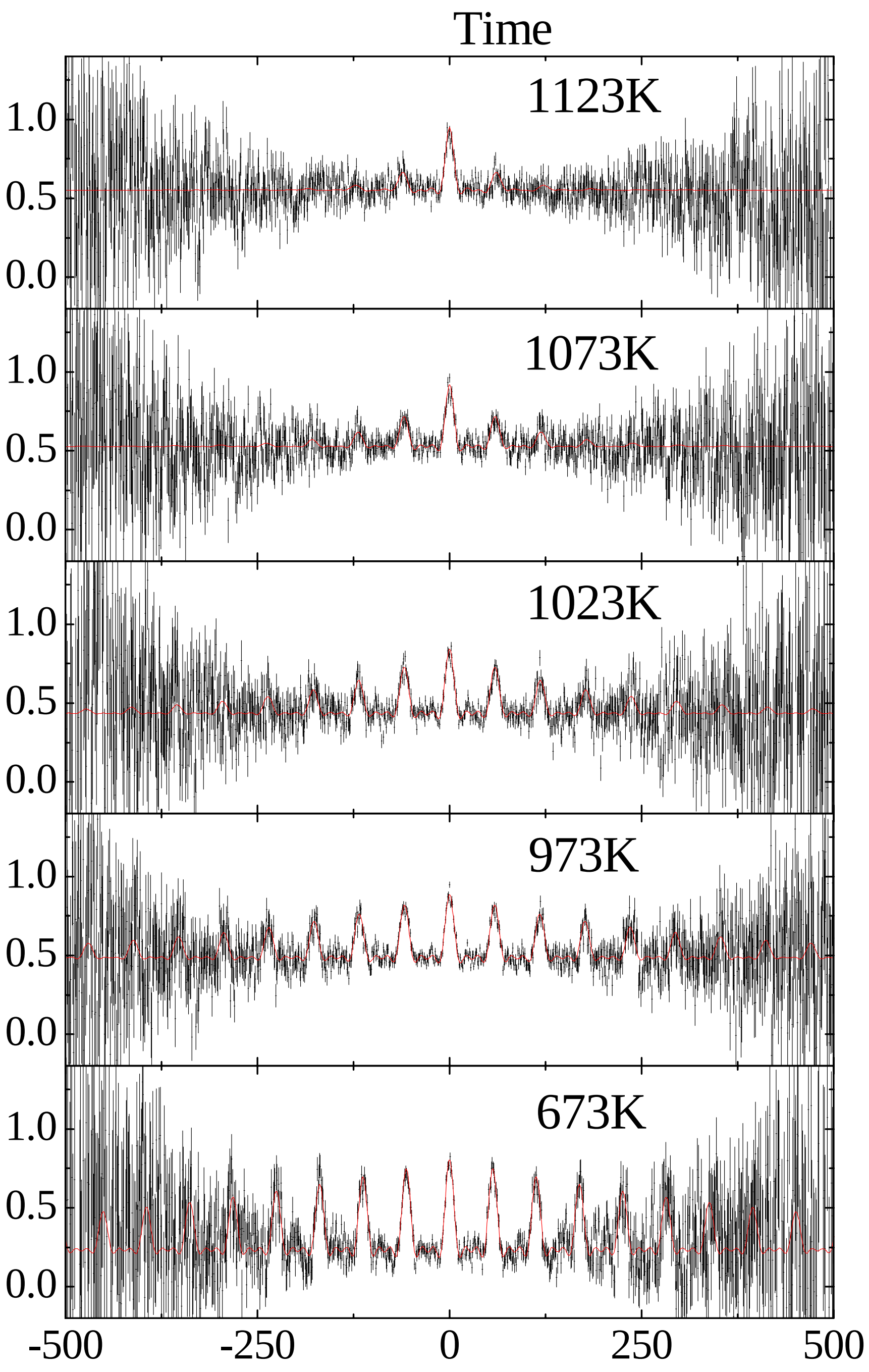

Fig. 2. PAC spectra for Pd-poorer $EuPd_3$. With increasing temperature, a shift in site preference can be seen from the high-frequency Pd-site to the low-frequency R-sites. Diffusional damping of the high-frequency signal increases with temperature.

Fig 3

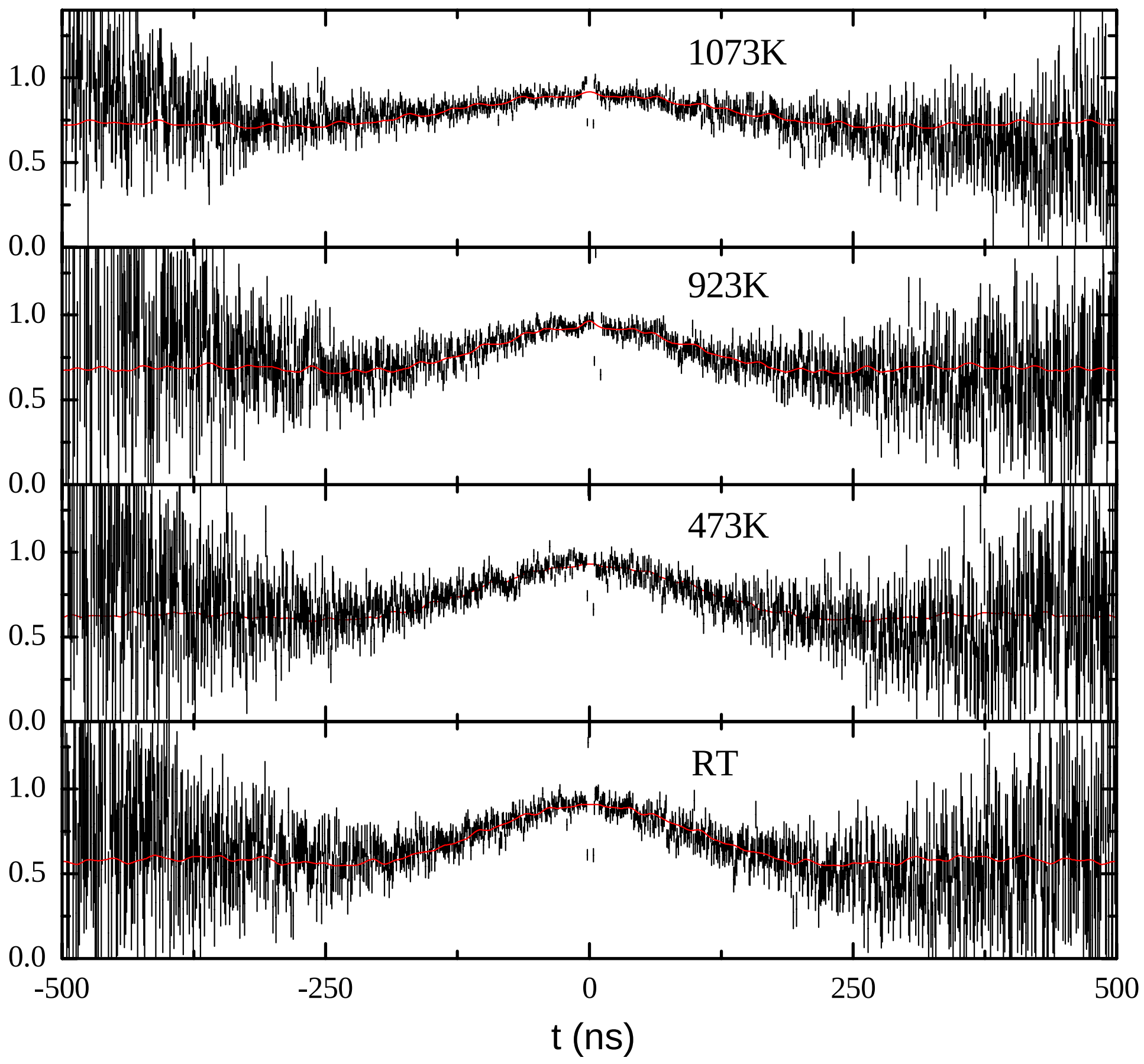

Fig 3. PAC spectra of $PrPd_3$ for the Pd-richer boundary composition. The low frequency interaction is attributed to indium probe atoms on the cubic Pr-sublattice.

Fig 4

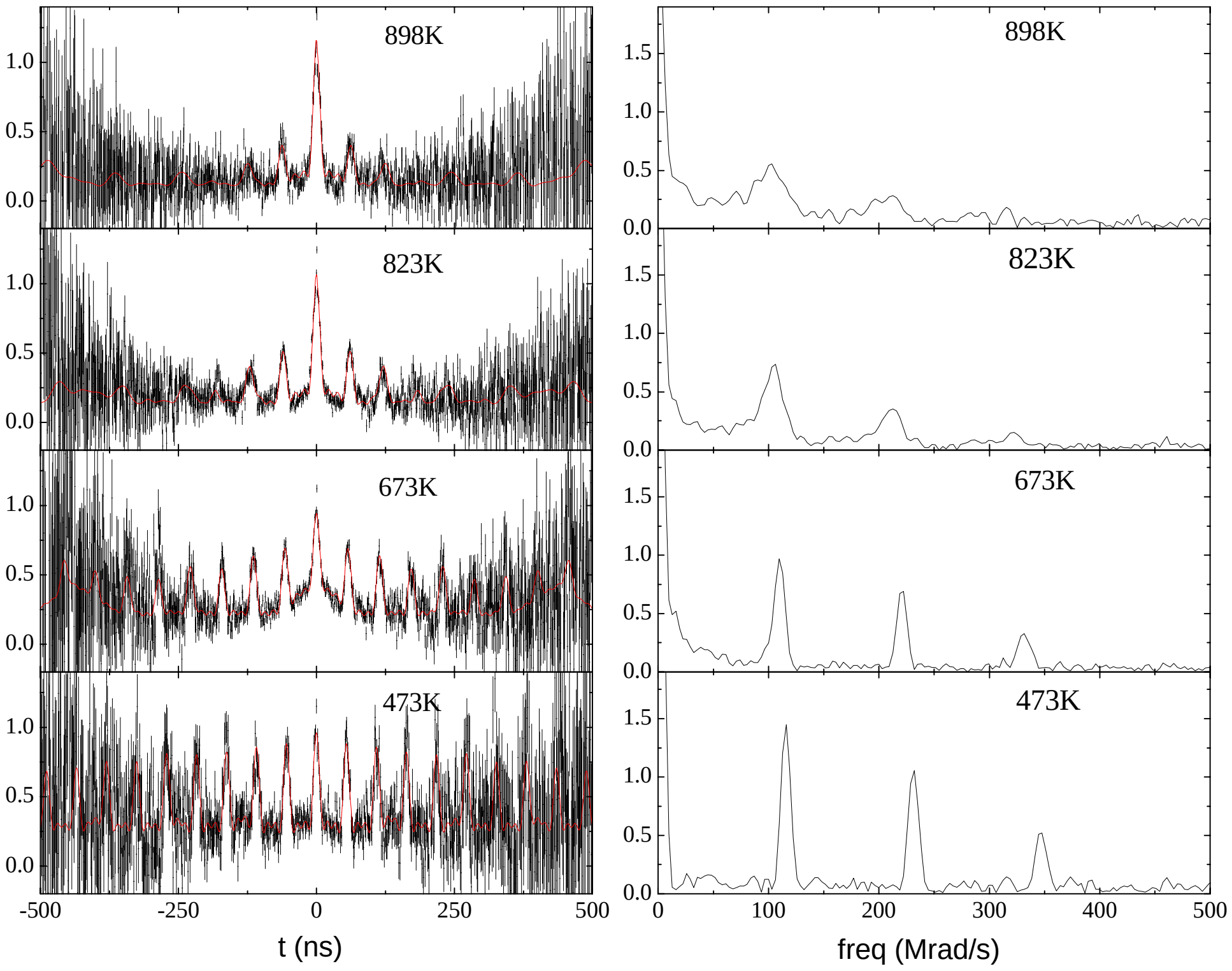

Fig 4. PAC spectra for $PrPd_3$ at the Pd-poorer boundary composition. Time domain spectra are on the left and fourier transforms on the right. Diffusional relaxation is observable as damping in the time-domain or frequency broadening in the frequency spectra.

Fig 5

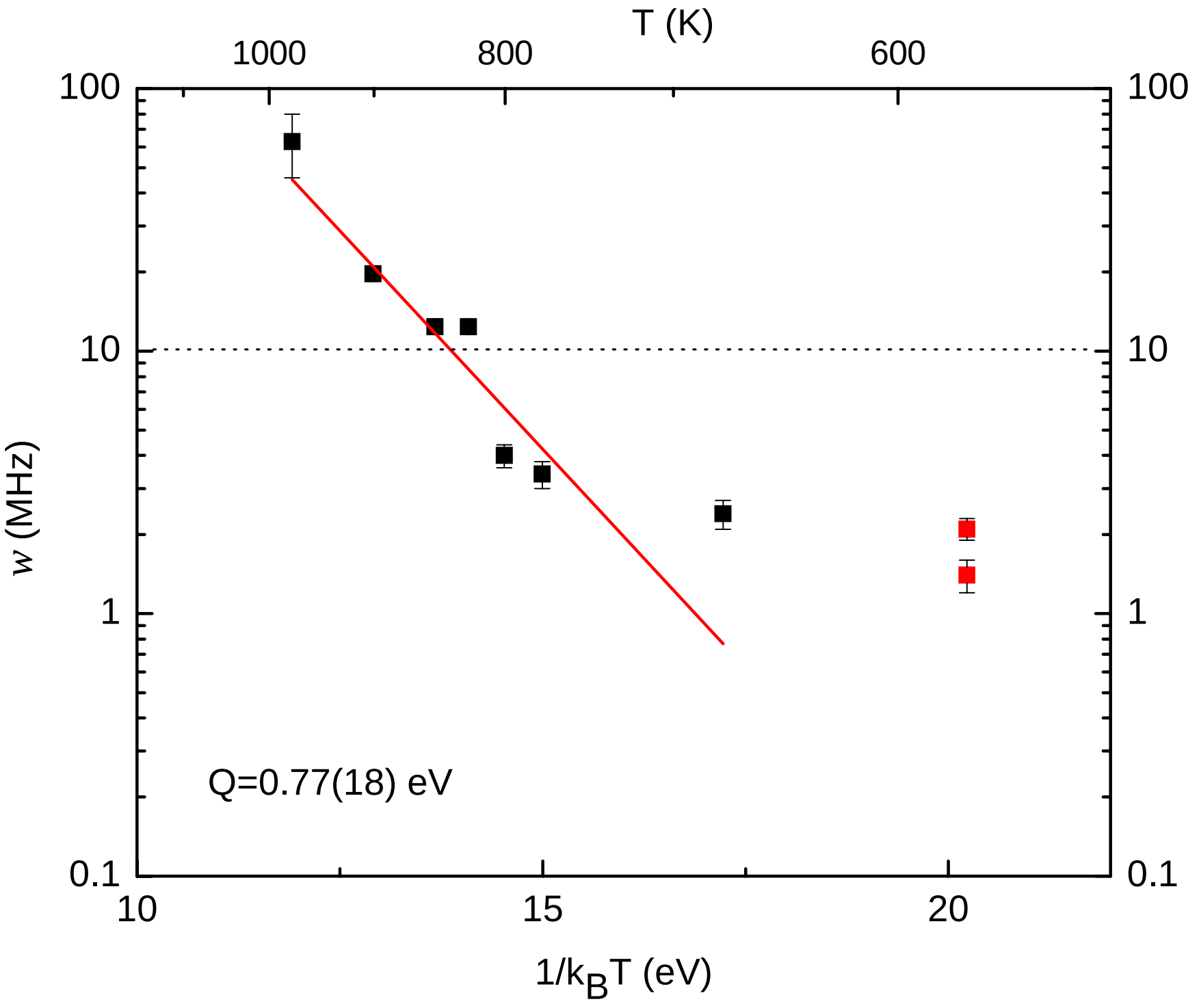

Fig. 5. Arrhenius plot of jump frequencies of indium probes in a sample of Pd-poorer $PrPd_3$. The two lowest temperature points were excluded from the fit shown by the drawn line.

Fig 6

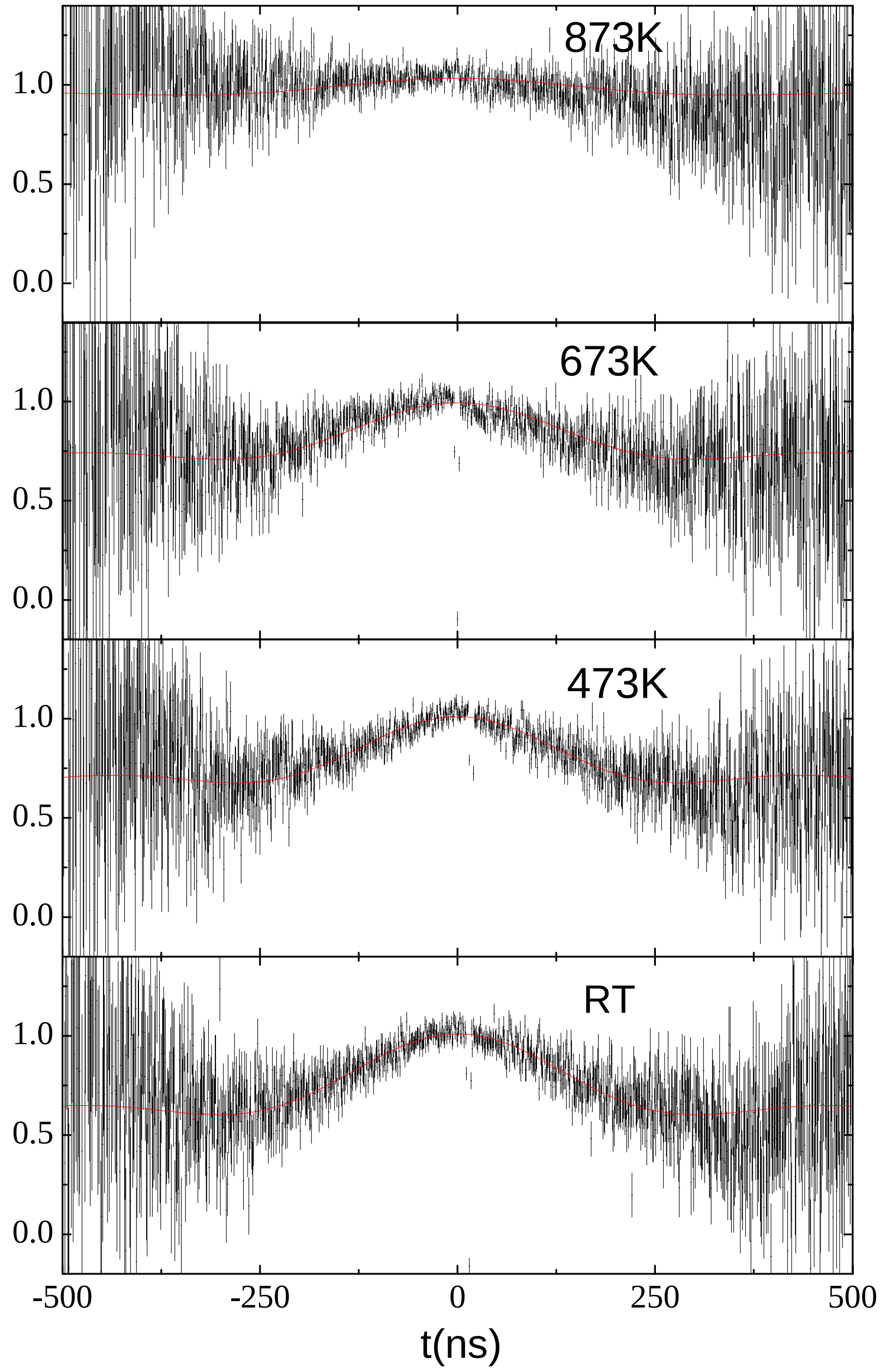

Fig 6. PAC spectra for Pd-poorer $LaPd_3$, exhibiting only the signal for indium solutes on R-sites.

Fig 7

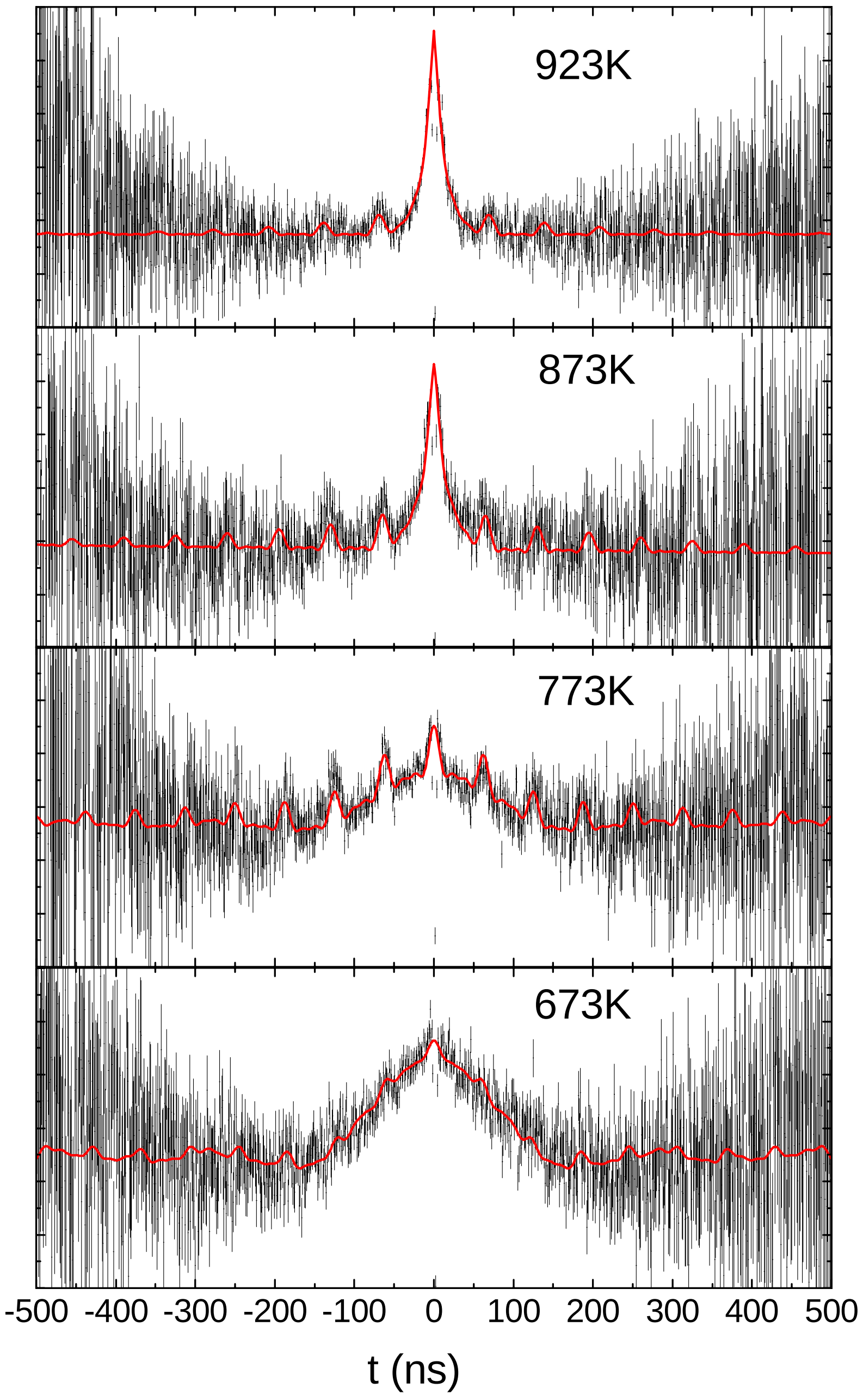

Fig 7. PAC spectra for Pd-poorer $CePd_3$ measured at the indicated temperature. With increasing temperature, there is a rapid shift in site preference from the Ce-site (low frequency) to the Pd-site (high frequency) and, at the same time, an increase in diffusional nuclear relaxation of the high-frequency signal. This combination of changes leads to complex spectra.

Fig 8

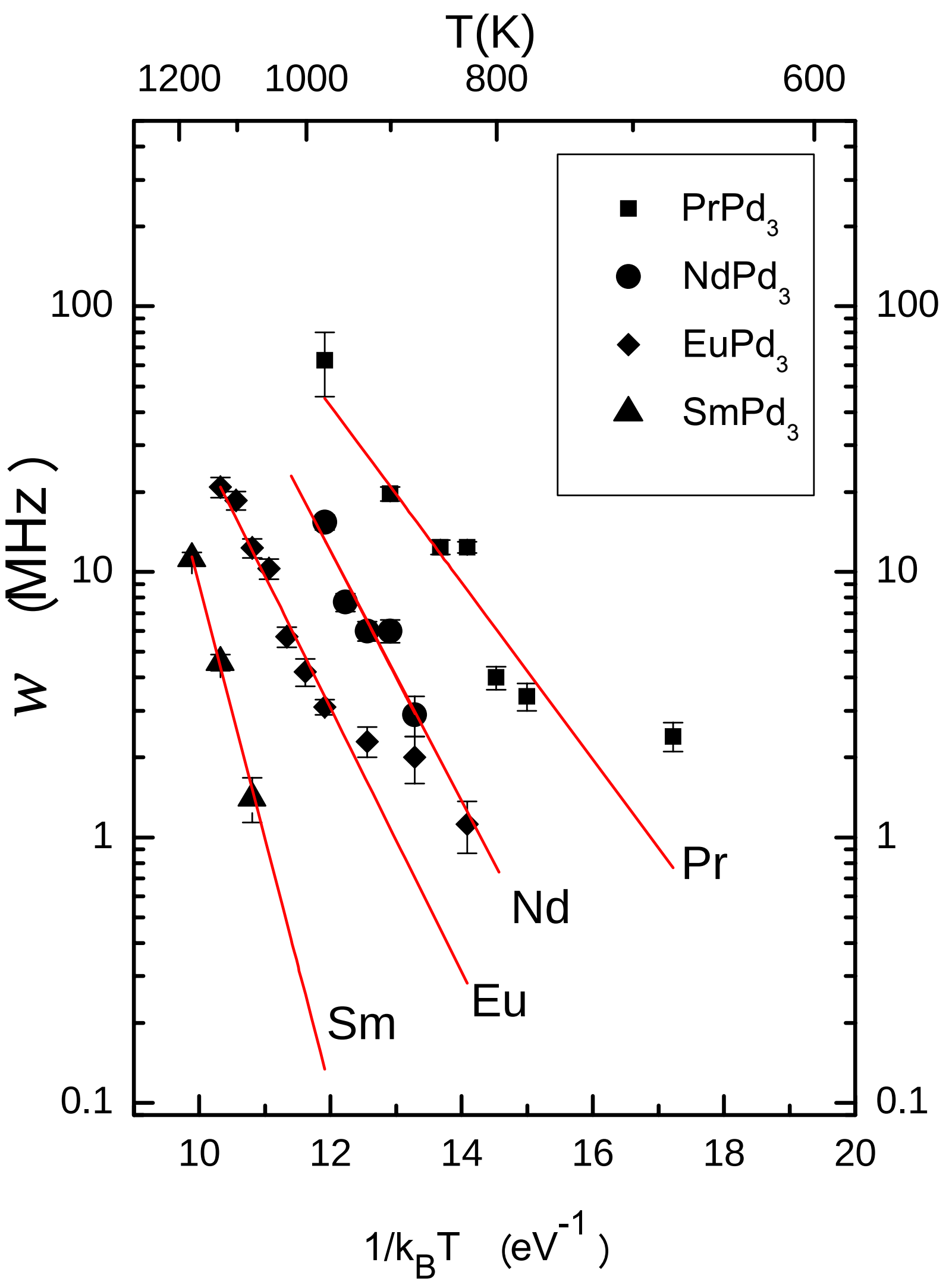

Fig 8. Arrhenius plots of jump frequencies of $^{111}$Cd probe atoms on the Pd-sublattices in the indicated $L1_2$ phases.

Fig 9

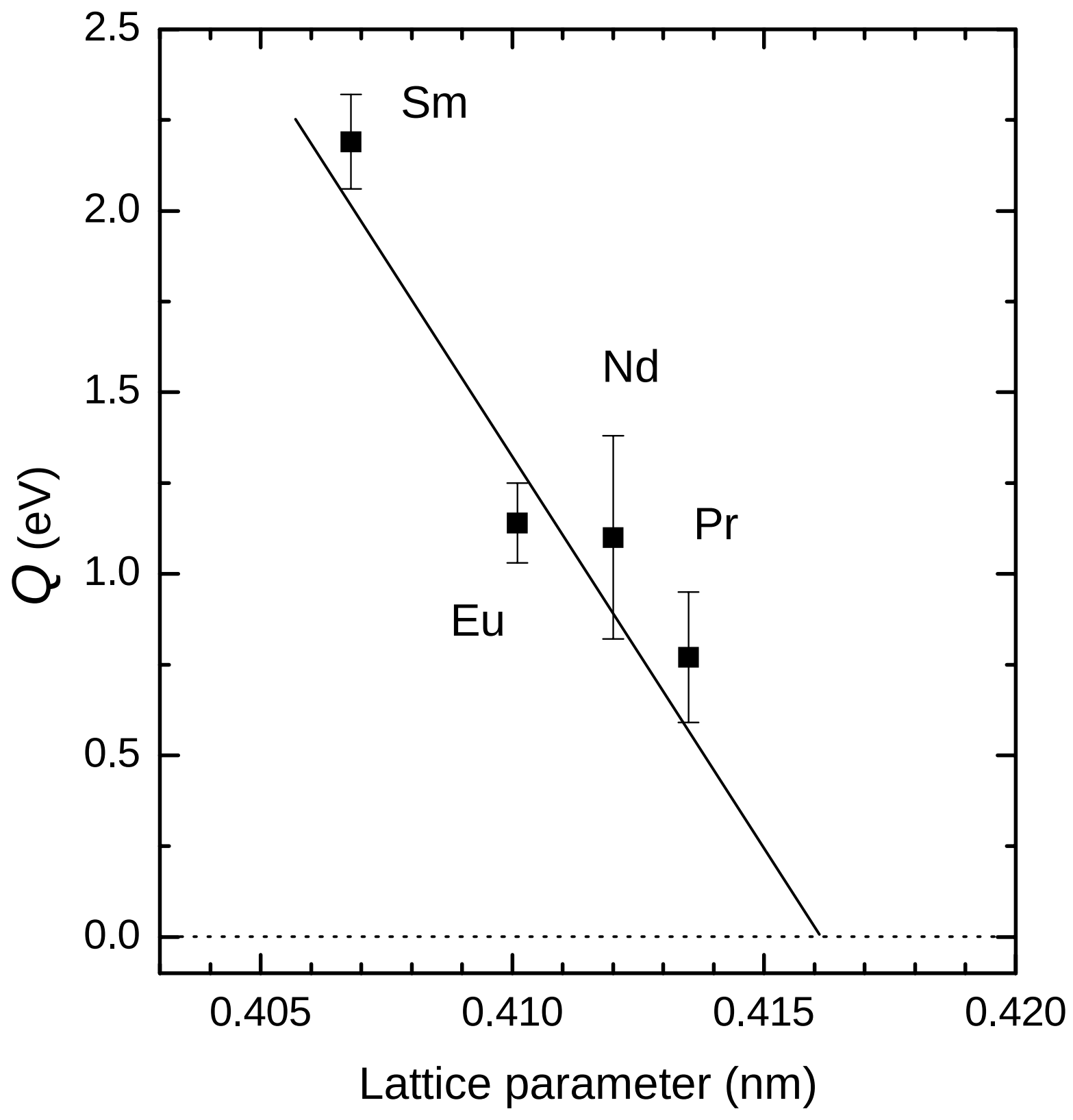

Fig 9. Activation enthalpies for diffusional jumps of Cd probe atoms on the Pd-sublattices of four $RPd_3$ phases plotted versus lattice parameter.

Fig 10

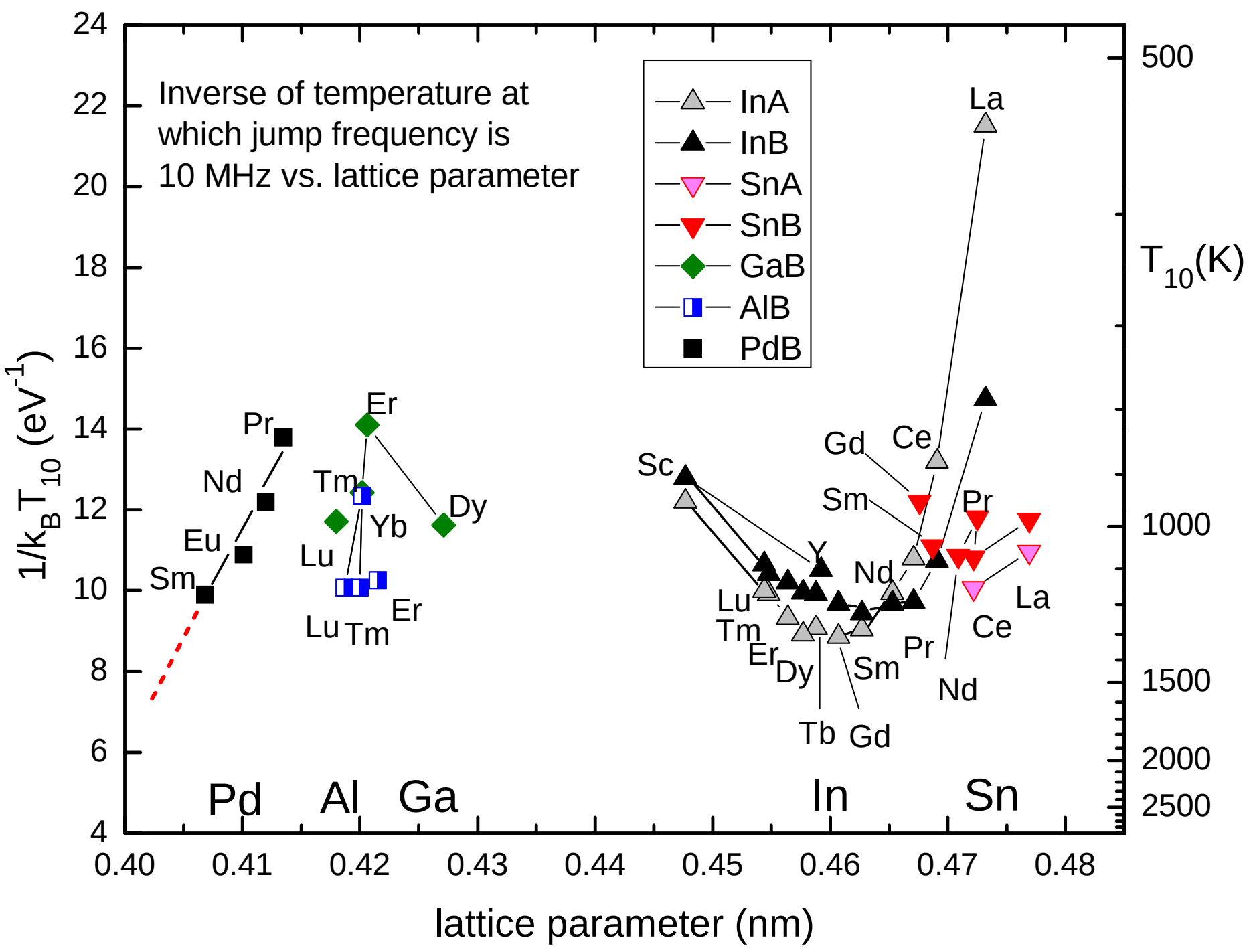

Fig. 10. Inverse temperatures at which jump frequencies of Cd probe atoms equal 10 MHz in $L1_2$ phases of rare-earth elements with In, Sn, Ga, Al and Pd. "A" and "B" designate R-poorer and R-richer boundary compositions.